\documentclass[fleqn,10pt]{wlscirep}
\usepackage[utf8]{inputenc}
\usepackage[T1]{fontenc}
\usepackage[normalem]{ulem}
\usepackage{multirow}
\usepackage{lineno}

\usepackage{tabularx}
\usepackage{float}
\usepackage{bm}
\usepackage{graphicx}
\usepackage{amsmath}
\usepackage{mathtools}
\usepackage{color}
\usepackage{mathrsfs}
\usepackage[export]{adjustbox}
\usepackage{bbold}

\newcommand{\blue}[1]{{\color{blue}#1}}

\definecolor{pacificb}{HTML}{1CA9C9}

\title{
Interaction of excitons with magnetic topological defects in 2D magnetic monolayers: localization and anomalous Hall effect \\
}

\author[1]{M.~Kazemi}
\author[2,*]{V.A.~Shahnazaryan}
\author[2,3]{Y.V.~Zhumagulov}
\author[1,2,4]{P.F.~Bessarab}
\author[1,2]{I.A.~Shelykh}

\affil[1]{Science Institute, University of Iceland, Dunhagi 3, IS-107, Reykjavik, Iceland}
\affil[2]{ITMO University, School of Physics and Engineering, St. Petersburg, 197101, Russia}
\affil[3]{University of Regensburg, Regensburg 93040, Germany}
\affil[4]{Department of Physics and Electrical Engineering, Linnaeus University, SE-39231 Kalmar, Sweden}
\affil[*]{Corresponding author}

\makeatletter
\renewcommand{\@maketitle}{%
{%
\thispagestyle{empty}%
\vskip-36pt%
{\raggedright\sffamily\bfseries\fontsize{20}{25}\selectfont \@title\par}%
\vskip10pt
{\raggedright\sffamily\fontsize{12}{16}\selectfont  \@author\par}
\vskip25pt%
}%
}%
\makeatother

\begin{document}

\flushbottom
\maketitle

\section*{ABSTRACT}

Novel 2D material $\text{CrI}_3$ reveals unique combination of  2D ferromagnetism and robust excitonic response.
We demonstrate that the possibility of the formation of magnetic topological defects, such as N\'eel skyrmions,  together with large 
excitonic Zeeman splitting, leads to giant scattering asymmetry, which is  
the necessary prerequisite for the excitonic anomalous Hall effect.
In addition, the diamagnetic effect breaks the inversion symmetry, and in certain cases can result in exciton localization on the skyrmion.
This enables the formation of magnetoexcitonic quantum dots with tunable parameters.

\thispagestyle{empty}

\section*{Introduction.}

Dimensionality has a dramatic impact on the physical properties of a system. Although we live in a three- dimensional (3D) world, the recent progress of technology allowed experimental realization of structures of lower dimensionality. In these systems the motion of a particle is restricted to one or two dimensions, which results in certain peculiarities of their physical behavior coming from dramatic enhancement of the role of quantum fluctuations. 

Two dimensional (2D) systems are of particular interest, as they lie between 3D systems where formation of Off Diagonal Long Range Order is possible at finite temperatures and 1D systems where fluctuations completely destroy long range order and lead to the exponential decay of quantum correlations in real space at any finite temperature. In 2D systems, the situation is more tricky: although long range correlations are destroyed, their decay at low temperature is much slower than in the 1D case. 
This is a characteristics of the Berezinskii-Kosterlitz-Thouless phase which is intimately connected with spontaneous creation of topological excitations. Their type depends on the nature of the system in question. In 2D magnets, they are skyrmions~\cite{Rossler2006,Kiselev2011,Romming2015}. For 2D superfluid systems, characteristic for geometries with excitons \cite{Butov2002,Astrakharchik2007,High2012,Lozovik2012,Fogler2014} or exciton polaritons \cite{Kasprzak2006,Balili2007,Amo2009}, such defects are vortex-antivortex pairs \cite{Lagoudakis2008,Tosi2012,Gao2018,Kwon2019}. 

Systems where several order parameters interact via particle-like entities exhibit particularly interesting, hybrid behavior. In multiferroics, magnetic skyrmions induce nontrivial dielectric order, which enables a mechanism for an electric-field control of skyrmions~\cite{seki2012}. In heterostructures combining chiral magnet and superconductor, co-existing magnetic skyrmions and superconducting vortices create a platform for nucleation and control of Mayorana fermions, a prerequisite for topological quantum computing~\cite{petrovic2021}.  

Recently, materials combining magnetic properties with robust excitonic response have been discovered~\cite{Burch2018}. These are 2D antiferromagnets  MPS$_3$ (M=Ni, Fe, Mn) \cite{Wildes2015,Kang2020,Ho2021,Birowska2021}, and ferromagnets CrX$_3$ (X=I, Br)~\cite{Huang2017,Zheng2018,Kashin2020,kim2019}. In seeming contradiction to the Mermin-Wagner theorem~\cite{Mermin1966}, the finite-temperature long-range magnetic order is enabled in these systems thanks to the magnetic anisotropy~\cite{Lado2017}. Even more intriguingly, application of electric field or stress breaks the structural inversion symmetry leading to the emergence of uncompensated anti-symmetric exchange between the magnetic atoms~\cite{Liu2018,Ghosh2019,Vishkayi2020} -- the Dzyaloshinskii-Moriya (DM) interaction -- that favors noncollinear alignment of magnetic moments. Interplay between magnetic anisotropy, Heisenberg exchange and DM interaction opens a possibility for complex magnetic order in 2D magnets, including the emergence of localized states with nontrivial topology such as skyrmions~\cite{Liu2018a,Behera2019}. 
\begin{figure}
   \centering
    \includegraphics[width = 0.70\linewidth]{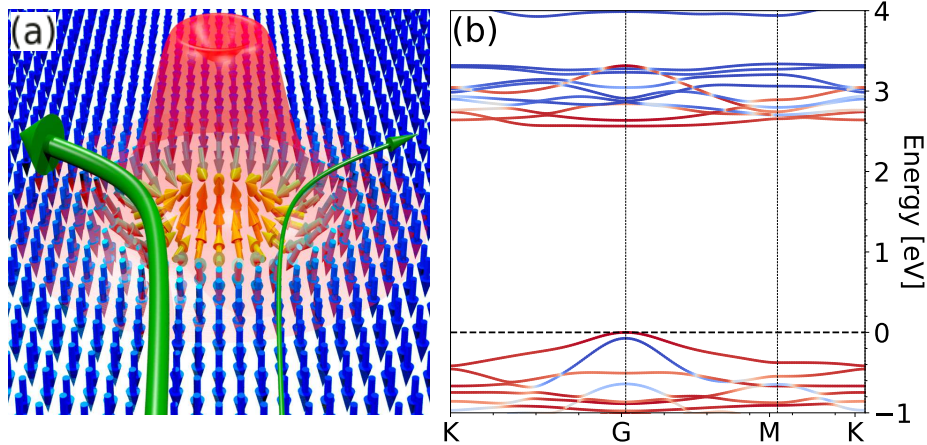}
    \caption{
    (a) Exciton scattering and localization on a skyrmion in CrI$_3$.
    The arrows indicate the calculated magnetization at each lattice site, with corresponding winding at the skyrmion area.
    The pair of green arrows illustrates the exciton scattering paths, with the arrow thickness being proportional to the scattering amplitude. 
    The red dome is the calculated exciton center-of-mass wave function, strongly localized on the skyrmion. 
    (b) DFT-based spin-resolved band structure of the CrI$_3$ monolayer. Colour codes the spin projection value. 
    }
    \label{fig:band_optical}
\end{figure}

The combination of huge optical oscillator strengths with giant Zeeman splitting in CrI$_3$ leads to various magneto-optical phenomena including giant Kerr response \cite{Huang2017}, magnetic circular dichroism \cite{Seyler2018}, onset of 2D magnetoplasmons \cite{Pervishko2020} and resonant inverse Faraday effect~\cite{Kudlis2021}. Moreover, potential emergence of magnetic skyrmions in 2D excitonic materials opens up new exciting physics and interesting applications.  

Here, we demonstrate the effects of scattering and localization of excitons on magnetic skyrmions in CrI$_3$, a representative 2D magnet, as illustrated in Fig.~\ref{fig:band_optical} (a). The scattering on a skyrmion is characterized by strong asymmetry, implying the possibility of Hall-like exciton transport \cite{Kozin2021}. The possibility to localize excitons leads to the formation of analogs of quantum dots with tunable properties, which can be 
used as polarization selective single photon emitters.

\section*{Results} 

\subsection*{Band structure and excitons}

Figure \ref{fig:band_optical} (b) shows the 
band structure of CrI$_3$ monolayer 
calculated using the density functional theory (DFT) (
see the \blue{Methods} section).
The 
bandgap 
is $E_g =  2.59$ eV, in agreement with previous calculations~\cite{Wu2019}.
We found the transition energy of the brightest low-lying exciton is $E_X =1.65$ eV. 
The corresponding exciton state is formed near direct bandgap at the crystallographic point G by the electrons mainly located in the
$m_c = 5.09m_0$, 
and the holes in the upper valence band with effective mass $m_v =  -0.59m_0$, with $m_0$ being the free electron mass. 
The exciton has a clearly defined spin structure since it is completely localised in the upper valence band and the three lower conduction bands, which have strictly determined spin direction. 
The most straightforward possible estimation of exciton Zeeman splitting is the difference between the energies of the lower conduction band and the fourth conduction band, which correspond to different spin directions, plus the energy difference between the upper conduction band and the second upper conduction band at the crystallographic point G. 
The resulting value of the Zeeman splitting is equivalent to 0.37 eV.

\subsection*{Magnetic skyrmions} 

Magnetic structure of CrI$_3$ can be described as a 2D honeycomb lattice of classical spins localized on Cr atoms. 
It is assumed that the system is subject to the external electric field breaking the structural inversion symmetry. 
As a result, in addition to the contributions due to the Heisenberg exchange, out-of-plane magnetic anisotropy and external magnetic field, the magnetic Hamiltonian of CrI$_3$ also includes the DM interaction energy~\cite{Behera2019}. Each contribution is characterized by an effective interaction parameter: $J$, $K$, and $D$ (see the \blue{Methods} section for the detailed description of the Hamiltonian). Magnitude of the interaction parameters is mostly taken from Ref.~\cite{Behera2019}, but some variation of the parameter values is also considered.
%

\begin{figure}
    \centering
    \includegraphics[width = 0.70\linewidth]{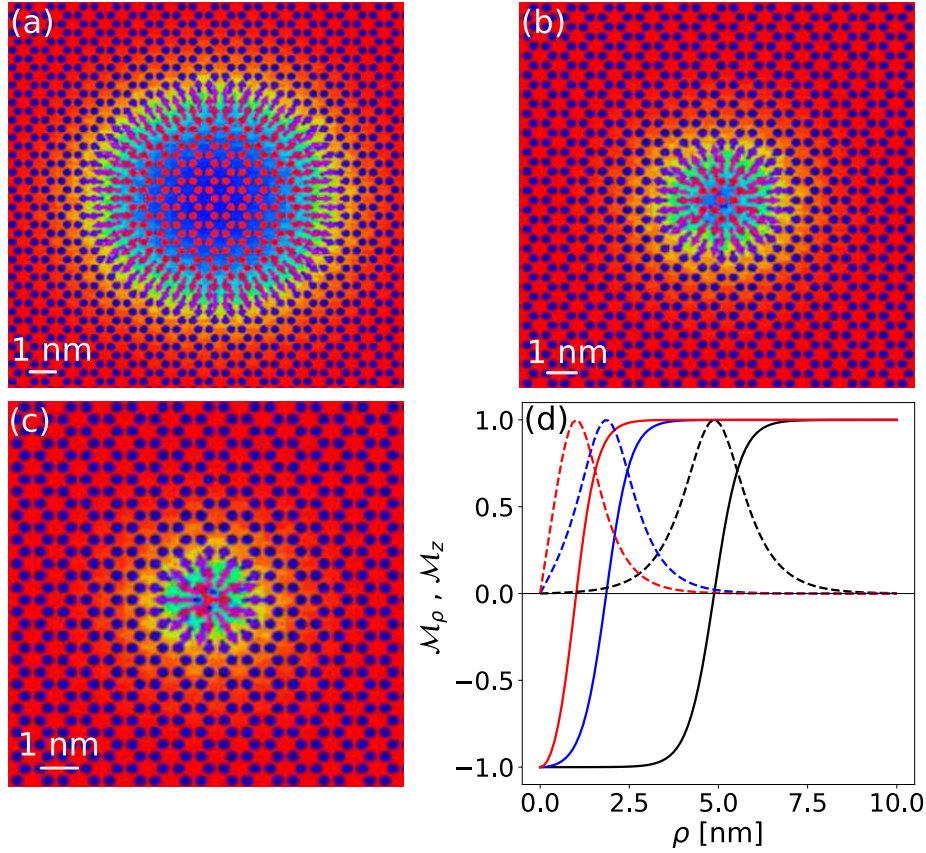}
    \caption{(a)-(c) Equilibrium magnetic configuration of N\'eel-type skyrmions in a 2D honeycomb lattice for various values of the magnetic interaction parameters. Color codes the out-of-plane component of magnetization. 
    (a) $D=0.59$~meV, $K=0.50$~meV, $J=1.26$~meV, $B=0$~T.
    (b) The same as in (a), but $D=0.54$~meV.
    (c) The same as in (a), but $B=0.8$~T.
    (d) The radial profiles of the magnetization vector components $\mathcal{M}_z$ (solid curves), and $\mathcal{M}_\rho$ (dashed curves)   corresponding to the configurations (a), (b), and (c) shown with black, blue, and red curves, respectively.
    }
    \label{fig:sk_prof}
\end{figure}

\begin{figure}
    \centering
    \includegraphics[width = 0.70\linewidth]{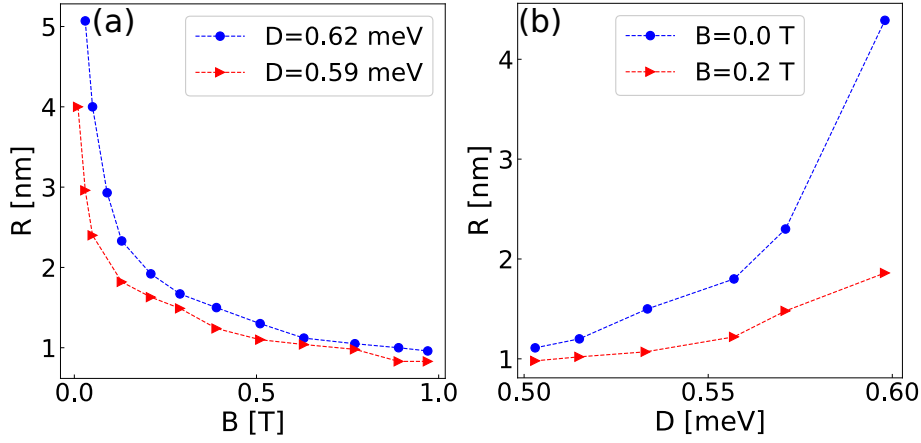}
    \caption{(a) The skyrmion radius versus the applied magnetic field for different values of the DM interaction parameter, as indicated in the legend.  
    (b) The skyrmion radius versus the DM interaction parameter for different values of the applied magnetic field, as indicated in the legend.
}
    \label{fig:sk_radius}
\end{figure}

Isolated skyrmion states are obtained via the energy minimization of the magnetic structure~\cite{ivanov2021} starting from a rough initial guess for the skyrmion profile.
Minimum-energy skyrmion state for $J=1.26$~meV, $D=0.59$~meV, $K=0.50$~meV, and zero magnetic field is shown in Fig.~\ref{fig:sk_prof}(a). 
How the skyrmion state changes with the DM interaction and external magnetic field is illustrated in Fig.~\ref{fig:sk_prof}(b) and ~\ref{fig:sk_prof}(c), respectively. 
Continuous representation of the skyrmion profiles is obtained by fitting the following ansatz to the simulation data~\cite{braun1994}: 

\begin{equation}
    \theta(\rho,c,w) = 2\arctan\left(\frac{\cosh(\frac{c}{w/2})}{\sinh(\frac{\rho}{w/2})}\right),
\end{equation}
where $\theta(\rho,c,w)$ is the polar angle of the magnetization at the distance $\rho$ from the skyrmion center and $c$, $w$ are the fit parameters. 
The normal and the radial components of the unit magnetization vector $\vec{\mathcal{M}}$ are defined as:

\begin{equation}
    \mathcal{M}_z = \cos\theta,\quad
    \mathcal{M}_\rho = \sin\theta.\label{eq:mrz}
\end{equation}
The skyrmion profiles corresponding to Fig.~\ref{fig:sk_prof}(a)-(c) are shown in Fig.~\ref{fig:sk_prof}(d). 
The skyrmion size can be obtained from the profiles as a radius $R$ of the $\mathcal{M}_z=0$ contours. 
The radius of skyrmions in CrI$_3$ demonstrates monotonic decrease (increase) with $B$ ($D$), see Fig.~\ref{fig:sk_radius}, which is consistent with the general theory of the skyrmion states~\cite{bogdanov1994a,bogdanov1994b}. 
The non-coplanar magnetic texture of a skyrmion creates an electromagnetic field~\cite{nagaosa2013} leading to the asymmetrical scattering and localization of excitons, as explained in what follows.

\subsection*{Exciton scattering}

Let us first consider the scattering of individual excitons on skyrmions.
We consider the center-of-mass dynamics of spinor exciton interacting with the magnetization field due to the skyrmion state. 
The corresponding Schr\"odinger equation has the form

\begin{align}
\label{eq:Schro}
\left[ -\frac{\hbar^2}{2\mu_X} \nabla^2 \mathbb{1}_2 + \hat{U}  \right] | \Psi \rangle = E_{\rm CM} | \Psi \rangle,
\end{align}
where $\mu_X$ is the exciton total mass, $\mathbb{1}_2$ is the $2\times2$ identity matrix, $\hat{U} = \lambda  \vec{\mathcal{M}}\vec{\sigma}$,
where $\lambda$ is the interaction energy, and $E_{\rm CM}$ is the exciton center of mass energy.
We restrict the treatment with the spin conserving exciton elastic scattering off the skyrmion. 
The general scattering of massive spinor particle on magnetic skyrmion is discussed in Refs. \cite{Denisov2016,Denisov2020}. 
In order to exclude the spin degree of freedom, we apply the adiabatic elimination approach \cite{Dalibard2011}, expanding the exciton state in the basis $|\Psi \rangle = \psi_1 |\chi_1\rangle + \psi_2 |\chi_2\rangle$.
Here $\chi_{1,2}$ are the eigenstates of skyrmion potential $\hat{U}$, corresponding to eigenvalues $ u_{1,2} = \mp \lambda$.
Assuming $\psi_2 = 0$, we project the Schr\"odinger equation to the state $|\chi_1\rangle$. 
It is of crucial importance, that in the effective spinless model, synthetic U(1) gauge field appears. 
It plays the role of an effective magnetic field and is responsible for giant asymmetry of the scattering as we will now see.

The resulting equation reads (see the \blue{Supplemental Material (SM)} for details):

\begin{align}
    \label{eq:psi1}
     \left[ \frac{1}{2\mu_X} (\hat{P} -e\vec{A})^2 + W - \lambda  \right] \psi_1 = E_{\rm CM} \psi_1,
\end{align}
where 

\begin{equation}
    \vec{A}  = 
    -\frac{\hbar}{e} \frac{1+ \mathcal{M}_z}{2\rho} \hat{\varphi},
\end{equation}
\begin{align}
    W & = \frac{\hbar^2 }{8\mu_X}  \left[ \left(\mathcal{M}_z \mathcal{M}_\rho' - \mathcal{M}_\rho \mathcal{M}_z'\right)^2
    + \mathcal{M}_\rho^2/ \rho^2  \right]
\end{align}
are vector and scalar gauge potentials, respectively.
Here $e$ denotes the unit charge, $\hat{\varphi}$ is the unit vector in the azimuthal direction, and the prime denotes the derivative with respect to 
$\rho$.
Interestingly, the gauge potentials do not depend on the interaction energy $\lambda$.
Consequently, the scattering cross-section in the adiabatic elimination approach is independent of  $\lambda$, as it only leads to rigid energy shift in Eq.~\eqref{eq:psi1}. 

\begin{figure}[t]
    \centering
    \includegraphics[width = 0.70\linewidth]{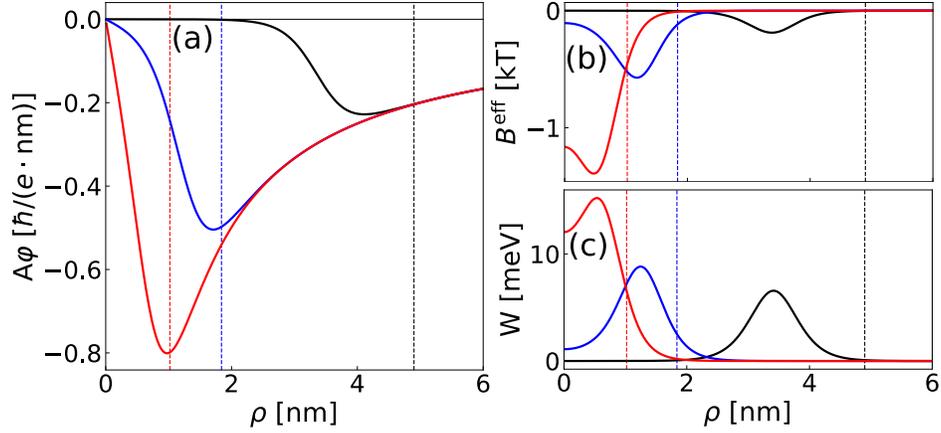}
    \caption{The radial dependence of (a) the gauge vector potential; (b)  effective magnetic field; (c) scalar potential for different profiles of skyrmion corresponding to Fig.~\ref{fig:sk_prof}. 
    The vertical lines denote the respective skyrmion radii. 
    The vector potential has only azimuthal component, which scales as $1/\rho$ outside the skyrmion region.
    The giant values of above 1000 T for the effective magnetic field are due to the nm size of the skyrmion and the rapid variation of vector potential inside.
    The scalar potential has a radial symmetry, and rapidly decays outside the skyrmion area.
    }
    \label{fig:pot}
\end{figure}

The presence of the vector potential $\vec{A}$ gives rise to an effective magnetic field $\vec{B}^{\rm eff} = \nabla \times \vec{A}$.
Due to the symmetry of the vector potential, the magnetic field is normal to the plane and reads

\begin{align}
    B_z^{\rm eff} = \frac{A_{\varphi}}{\rho} +  \frac{\partial A_{\varphi}}{\partial \rho}.
\end{align}
The radial dependence of the vector potential is presented in Fig.~\ref{fig:pot} (a).
Notably, the maximal absolute value of the vector potential decreases versus the skyrmion size. 
Conversely, the maxima appears at larger distance from the skyrmion center.
Outside the skyrmion region,  the vector potential demonstrates a $1/\rho$ scaling. 
Figure~\ref{fig:pot} (b) shows the emergent effective magnetic field due to skyrmion.
The giant values of the field are explained by the rapid variation of vector potential at sub-nm scale, driven by the nature of the skyrmion magnetization texture. 
We mention that for a skyrmion having the size of $\sim 16$ nm due to topological winding of the texture an effective magnetic field of $B^{\rm eff} \approx 13$ T was reported earlier \cite{Ritz2013}. 

The radial dependence of scalar potential is shown in Fig.~\ref{fig:pot} (c), demonstrating a good localization within the skyrmion. Notably, it has a local minima at $\rho=0$, which can lead to the formation of the excitonic quasibound states and scattering resonances. However, characteristic energies of these resonances correspond to the wavevectors of the excitons lying far outside the light cone, which can not be directly created optically, and are thus not analyzed in this work.
\begin{figure}
    \centering
    \includegraphics[width = 0.70\linewidth]{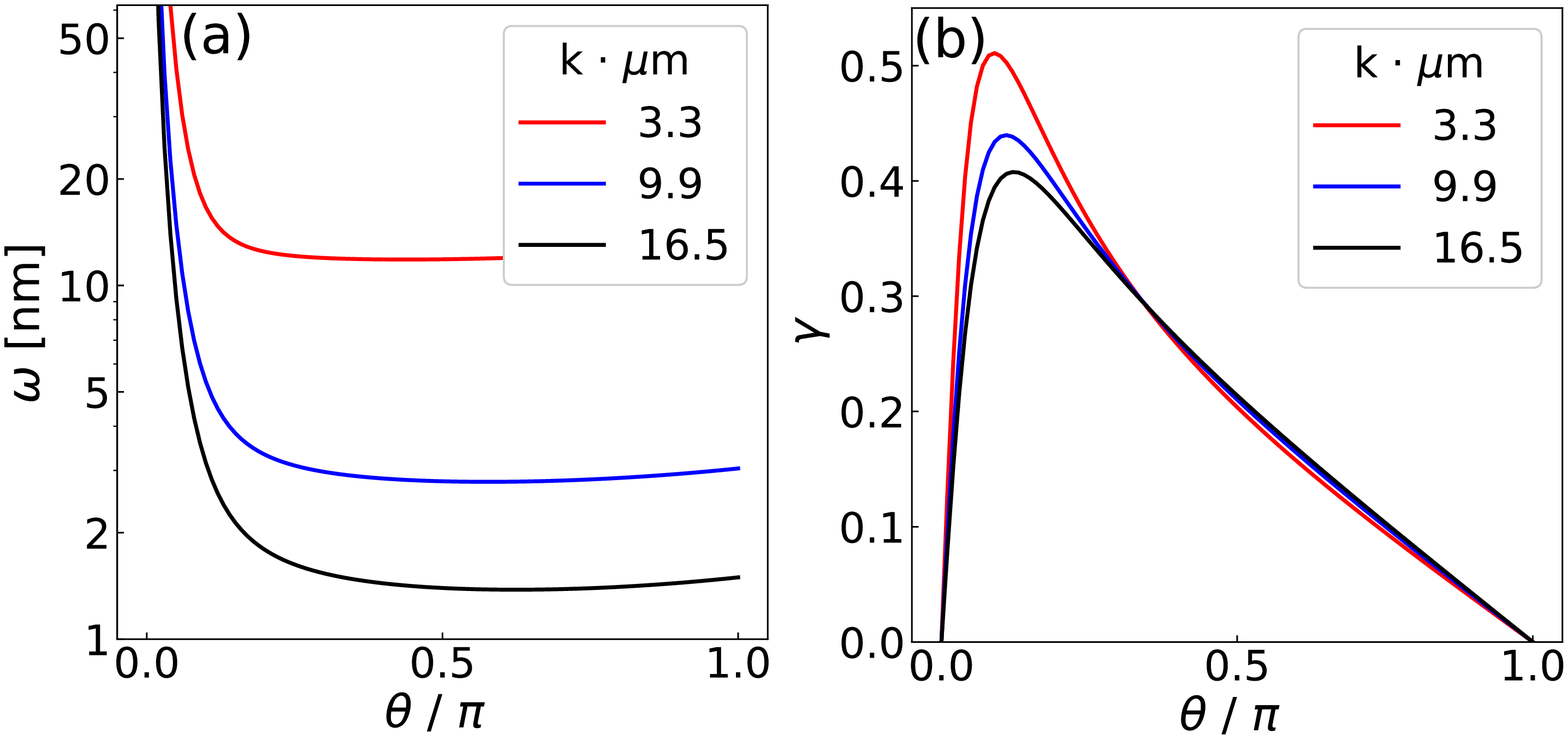}
    \caption{ 
    The cross-section of exciton scattering on skyrmion. 
    (a) The angular dependence of amplitude for several values of the wave vector, as indicated in the legend.
    The scattering amplitude monotonically decays with wave vector, and varies weakly with the scattering angle,  with the rapid divergence at $\theta \rightarrow 0$ region.
    (b) The scattering asymmetry 
    $\gamma = |\omega(k,\theta)-\omega(k,-\theta)|/\left(\omega(k,\theta)+\omega(k,-\theta) \right)$ versus  the scattering angle for several values of the wave vector, as indicated in the legend.
    The giant asymmetry at small scattering angles is accompanied by moderate dependence on the wave vector. 
    }
    \label{fig:scat}
\end{figure}
\begin{figure}
    \centering
    \includegraphics[width = 0.70\linewidth]{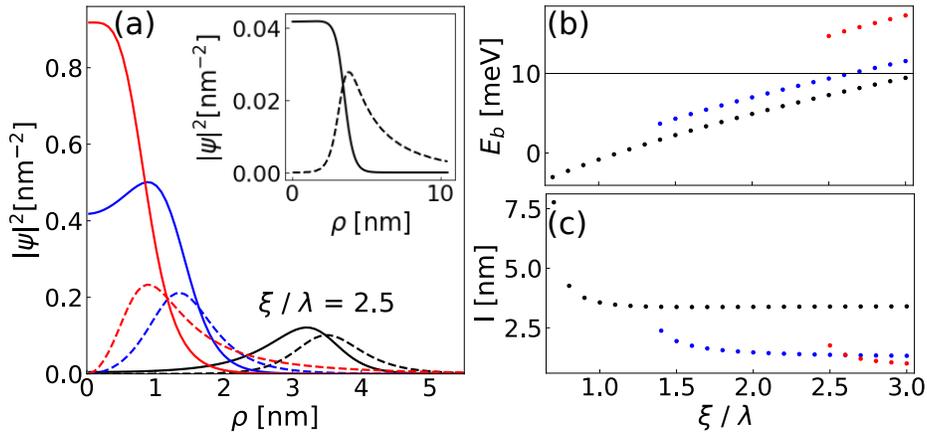}
    \caption{ 
    (a) The center-of-mass wave function of exciton localized on skyrmion.
    The colors correspond to different profiles of skyrmion shown in Fig. \ref{fig:sk_prof}.
    The solid and dashed curves correspond to spin-up and spin-down components of the wave function. 
    Here the diamagnetic ratio is $\xi / \lambda = 2.5$, and the inset illustrates the case of a large skyrmion, with moderate diamagnetic ratio $\xi / \lambda = 0.7$.
    (b) The binding energy, and (c) the radius of exciton localization on skyrmion versus the diamagnetic ratio. For smaller skyrmions, the localized states emerge only at very large values of diamagnetic shift. The thin line in panel (b) shows the interaction energy $\lambda$. 
    }
    \label{fig:local}
\end{figure}

The differential cross-section can be expanded up to third order in skyrmion potential strength \cite{Sinitsyn2007}:

\begin{align}
    \omega(\vec{k},\vec{k}') \approx 
    \omega^{(2)}(\vec{k},\vec{k}') 
    + \omega_s^{(3)}(\vec{k},\vec{k}')
    +\omega_a^{(3)}(\vec{k},\vec{k}'),
\end{align}
where the lowest order term $\omega^{(2)}(\vec{k},\vec{k}')$ is symmetric. 
The third order term contains an irrelevant symmetric correction $\omega_s^{(3)}(\vec{k},\vec{k}')$ and an asymmetric contribution $\omega_a^{(3)}(\vec{k},\vec{k}')$ (see the \blue{Methods}).
The cross-section of exciton scattering  is presented in Fig.~\ref{fig:scat}. 
The exciton wave vector has upper limit defined by optical excitation range $k_{\rm max} = n E_X / (\hbar c)$. 
Here $n\approx2$ is the refractive index of surrounding media \cite{Wu2019}, $c$ is the speed of light. Together with excitonic resonance $E_X = 1.65$ eV, this yields 
$k_{\rm max} \approx 16.5$ $\mu m^{-1}$,  
and the exciton total mass is $\mu_X =m_e + |m_h| \approx 5.6 m_0$.
Figure~\ref{fig:scat} (a) illustrates the scattering amplitude versus the scattering angle for several values of the wave vector.
The divergent character of scattering at $\theta \rightarrow 0$ limit is due to the Coulomb-like tale of the vector potential outside the skyrmion region \cite{Aharonov1959}. 
For the characterization of scattering asymmetry we introduce the asymmetry degree

\begin{equation}
\gamma = \frac{|\omega(k,\theta)-\omega(k,-\theta)|}{\omega(k,\theta)+\omega(k,-\theta)}    
\end{equation}
This quantity is illustrated in Fig.~\ref{fig:scat} (b). 
One clearly sees, that scattering is highly asymmetric ($\gamma$ can reach values of 0.5), and the asymmetry rate has a moderate dependence on scattering wave vector.
On the contrary, the angular dependence is characterized by a sharp maxima at $\theta \approx \pi/12$, position of which weakly depends on the exciton wave vector. 
As some of us demonstrated earlier, asymmetric scattering can lead to the excitonic anomalous Hall effect \cite{Kozin2021}.

\subsection*{Exciton localization}

In the regime of large magnetic fields the diamagnetic term has a strong impact on the excitonic spectrum. 
Given by ferromagnetic nature of considered material, in Eq.~\eqref{eq:Schro} we account for quadratic terms in magnetic field, which can be phenomenologically included in our model:  

\begin{equation}
\hat{H}_{\rm dia} = \xi  \mathcal{M}_z^2 \mathbb{1}_2,   
\end{equation}
where $\xi$ is the diamagnetic shift energy, which is taken as a phenomenological parameter.
At sufficiently large values of $\xi$ the presence of diamagnetic term can result in exciton localization at skyrmion.
We seek for the $s-$state solution in the form
$    \Psi = (  \psi_+ , \psi_- e^{i\varphi} )$. 
The radial profiles of the components of exciton wave function are plotted in Fig.~\ref{fig:local} (a). 
Quite naturally, for smaller skyrmion the area of exciton localization is more compact. 
Further, with the enhancement of diamagnetic term the binding energy  increases, as shown in Fig.~\ref{fig:local} (b).
For the skyrmion of smaller radius the localized state appears at larger values of diamagnetic shift.
Figure~\ref{fig:local} (c) 
illustrates the effective radius of localization $l = \langle \Psi |\rho|\Psi \rangle$.
The saturation of localization radius at elevated values of diamagnetic shift is explained by the repulsion of exciton wave function components towards the skyrmion edge, c.f. the black curves in Fig.~\ref{fig:local} (a) and the inset therein.

\section*{Conclusions and Discussion}

In conclusion, we developed a theory addressing the co-existence of excitons and magnetic skyrmions in a 2D material. Within the theory, the band structure and exciton states are calculated using DFT 
and the magnetic properties including the properties of skyrmion states are obtained via atomistic spin simulations. We applied the theory to the celebrated CrI$_3$ system, yet the phenomena we predicted should be relevant for a class of 2D magnetic semiconductors with the possibility of broken structural inversion symmetry necessary for the emergence of uncompensated DM interaction. 

We revealed 
two 
scenarios of exciton-skyrmion interaction. 
The first scenario implies a spin-conserving exciton elastic scattering off the skyrmion. 
Here we apply an adiabatic elimination approach to discard the spin degree of freedom. 
Similar to the scattering of massive spinor paricle \cite{Denisov2016}, the exciton scattering in spinless domain is characterized by giant asymmetry, provided by emerging synthetic U(1) gauge field.
Such asymmetric scattering is a necessary prerequisite of the excitonic anomalous Hall effect \cite{Kozin2021}.
The second scenario accounts for the impact of diamagnetic effect on the exciton-skyrmion interaction.
The diamagnetic term breaks the inversion symmetry, which for sufficiently large values of diamagnetic energy shift leads to an efficient localization of spinful exciton on the skyrmion.
This opens a way for creation of tunable analogs of magnetic excitonic quantum dots, which can find their applications in the domain of quantum optics, for example as single photon emitters with tailored properties. Additionally, the exciton localization enables optical detection of skyrmions, which is particularly important for antiferromagnetic systems.

In our work, we focus on excitonic response to the presence of magnetic skyrmions, 
but the opposite effect is also feasible thus enabling resonant optical control of skyrmions. Generalization of the theory to describe the dynamical response of magnetic skyrmions should be straightforward~\cite{Kudlis2021}. This will make it possible to simulate various phenomena associated with coupled exciton-skyrmion dynamics. 

\section*{Methods}

\subsection*{Calculation of band structure and exciton transitions}

The band structure of a single-layered CrI3 is calculated using the density functional theory (DFT) approach implemented in the GPAW \cite{Mortensen2005,Enkovaara2010,Yan2011} code with LDA exchange-correlation functional.  
The lattice constant is taken as $a_0 = 6.69$ $\AA$ \cite{Wu2019} and the vacuum distance is 16 $\AA$. 
The spin-orbit interaction is taken into account using the first-order perturbation theory \cite{Olsen2016}.
The ground states calculation is done using the plane-wave basis with the $6\times6\times1$ k-space grid with the plane wave cutoff equal to 600 eV. Exciton states are calculated using {\it ab initio}
Bethe-Salpeter equation (BSE) method implemented in GPAW code \cite{Huser2013,Olsen2021} on the $6\times6\times1$ $k$-space grid with 16 valence and 14 conduction bands with 50 eV plane wave cutoff. 
In order to correct the DFT bandgap, we apply a scissor operator equal to 1.77 eV \cite{Levine1989}. 

\subsection*{Simulated magnetic subsystem} 

Magnetic subsystem of CrI$_3$ is modeled as a single monolayer of classical vectors localized on vertices of a honecomb lattice.The magnetic energy of the system reads:
\begin{equation}
\begin{split}
E=-\frac{J}{2}\sum_{<i,j>}\vec{m}_i\cdot\vec{m}_j-\frac{D}{2}\sum_{<i,j>}\vec d_{ij} \cdot \left[\vec{m}_i\times\vec{m}_j\right] \\
- K \sum_i\left(\vec{m}_i\cdot\vec{e}_z\right)^2-\mu\sum_i \vec{B}\cdot \vec{m}_i
\end{split}
\end{equation}
Here, $\vec{m}_i$ is a unit vector in the direction of $i$th magnetic moment with magnitude $\mu=3$~Bohr magnetons, $\vec{B}$ is the external magnetic field, and $d_{ij}=\vec r_{ij}\times \vec{e}_z/|\vec r_{ij}|$ is the DM unit vector with $\vec{e}_z$ and $\vec r_{ij}$ being the unit vector along the monolayer normal and the vector  pointing from site $i$ to site $j$, respectively \cite{Yang2015}. The pairwise interactions are between the nearest neighbors only. Both the magnetic field and the anisotropy axis are along $\vec{e}_z$. The size of the computational domain is chosen to be $50\times50$ unit cells. Periodic boundary conditions are applied so as to model the extended two-dimensional systems. 

\subsection*{Exciton scattering}

The differential cross-section of elastic scattering reads \cite{Denisov2016,Adhikari1986}:

\begin{align}
    \omega(\vec{k},\vec{k}')
    = \frac{\mu_X^2}{ 2\pi \hbar^4}\frac{|T(\vec{k},\vec{k}')|^2}{k} .
\end{align}
Here $\vec{k}$, $\vec{k}'$ correspond to the center of mass wave vectors of incident and scattered excitons, respectively, with $k= |\vec{k}| = |\vec{k}'|$.
The scattering $T$-matrix can be expanded as

\begin{align}
    T(\vec{k},\vec{k}') \approx V(\vec{k},\vec{k}') +  \sum_{\vec{k}''} \frac{ V(\vec{k}',\vec{k}'') V(\vec{k}'',\vec{k}) }{\epsilon_k - \epsilon_{k''}+ i\eta},
    \label{eq:Tmatr}
\end{align}
where $\epsilon_k = \hbar^2 k^2 /(2\mu_X)$.
The matrix element $V(\vec{k}, \vec{k}')$ within the first Born approximation has the form (see the \blue{SM} for the derivation):

\begin{align}
    V(\vec{k}',\vec{k}) &=  
    \frac{1}{2\pi} \int\limits_0^\infty J_0 (|\Delta \vec{k}| \rho)  \left(\frac{e^2 \vec{A}^2 }{2\mu_X}  + W\right) \rho {\rm d} \rho  \notag \\ 
    &- \frac{i}{\pi} \frac{e\hbar}{2\mu_X} \frac{[\vec{k}' \times \vec{k}]_z}{|\Delta \vec{k}| }  
    \int\limits_0^\infty J_1(|\Delta \vec{k}| \rho) A(\rho)  \rho {\rm d} \rho ,
\end{align}
where $\Delta \vec{k} = \vec{k}'-\vec{k}$.
Thus, the differential cross-section can be expanded up to third order in skyrmion potential strength \cite{Sinitsyn2007}:

\begin{align}
    \omega(\vec{k},\vec{k}') \approx 
    \omega^{(2)}(\vec{k},\vec{k}') 
    + \omega_s^{(3)}(\vec{k},\vec{k}')
    +\omega_a^{(3)}(\vec{k},\vec{k}'),
\end{align}
where the lowest order term 

\begin{align}
    \omega^{(2)}(\vec{k},\vec{k}') 
    = \frac{\mu_X^2}{ 2\pi \hbar^4}\frac{|V(\vec{k},\vec{k}')|^2}{k} 
\end{align}
is symmetric. 
The third order terms read: 

\begin{align}
    \omega_s^{(3)}(\vec{k},\vec{k}') 
     = \frac{\mu_X^2}{ 2\pi \hbar^4 k} 
     2 \mathcal{P} \left\{ {\rm Re} \left[ V(\vec{k},\vec{k}') \sum_{\vec{k}''} \frac{ V(\vec{k}',\vec{k}'') V(\vec{k}'',\vec{k}) }{\epsilon_k - \epsilon_{k''}} \right] \right\} ,
\end{align}
\begin{align}
    \omega_a^{(3)}(\vec{k},\vec{k}') 
    & = -\frac{\mu_X^2}{  \hbar^4 k } 
    {\rm Im} \left[ V(\vec{k},\vec{k}') \sum_{\vec{k}''}  V(\vec{k}',\vec{k}'') V(\vec{k}'',\vec{k}) \delta(\epsilon_k - \epsilon_{k''} ) \right] ,
    \label{eq:omegaa}
\end{align}
where $\mathcal{P}$ denotes the principal value integration.
The details of evaluation of Eq.~\eqref{eq:omegaa} are presented in \blue{SM}.

\section*{Acknowledgments}

We thank Prof. I.V. Iorsh for valuable discussions. The work was supported by the Icelandic Research Fund Grants 163082-051 and 217750, the University of Iceland Research Fund (Grant No. 15673), and the Swedish Research Council (Grant No. 2020-05110). V. S. acknowledges the University of Iceland for the hospitality.

\section*{Author contributions statement}
Y.V.Z. performed the DFT calculations of band structure and exciton states.
M.K. and P.F.B. calculated the magnetic skyrmion states.
M.K. and V.A.S. calculated the exciton scattering and localization on skyrmion. 
I.A.S. formulated the original idea and supervised the work.
All authors extensively discussed the results and  participated in editing of the manuscript.

\section*{Additional information}
The authors declare no competing interests.


\begin{thebibliography}{99}

\bibitem{Rossler2006} U. K. R{\"o}{\ss}ler, A. N. Bogdanov, and C. Pfleiderer, 
Spontaneous skyrmion ground states in magnetic metals, 
\href{https://www.nature.com/articles/nature05056}{Nature  {\bf 442}, 797 (2006)}.

\bibitem{Kiselev2011} N. S. Kiselev, A. N. Bogdanov, R. Sch\"afer and U. K. R\"o{\ss}ler, 
Chiral skyrmions in thin magnetic films: new objects for magnetic storage technologies?, 
\href{https://iopscience.iop.org/article/10.1088/0022-3727/44/39/392001}{J. Phys. D: Appl. Phys. {\bf 44,} 392001 (2011)}.

\bibitem{Romming2015} N. Romming, A. Kubetzka, C. Hanneken, K. von Bergmann and R. Wiesendanger, 
Field-dependent size and shape of single magnetic skyrmions, 
\href{https://journals.aps.org/prl/abstract/10.1103/PhysRevLett.114.177203}{Phys. Rev. Lett. {\bf 114,} 177203 (2015)}.

\bibitem{Butov2002} L. V. Butov, A. C. Gossard, and D. S. Chemla,
Macroscopically ordered state in an exciton system,
\href{https://www.nature.com/articles/nature00943}{Nature {\bf 418}, 751 (2002)}.

\bibitem{Astrakharchik2007} 
G. E. Astrakharchik, J. Boronat, I. L. Kurbakov, and Y. E. Lozovik,
Quantum Phase Transition in a Two-Dimensional System of Dipoles,
\href{https://journals.aps.org/prl/abstract/10.1103/PhysRevLett.98.060405}{Phys. Rev. Lett. {\bf 98}, 060405 (2007)}.

\bibitem{High2012} 
A. A. High, J. R. Leonard, A. T. Hammack, M. M. Fogler, L. V. Butov, A. V. Kavokin, K. L. Campman, and A. C. Gossard,
Spontaneous coherence in a cold exciton gas,
\href{https://www.nature.com/articles/nature10903}{Nature {\bf 483}, 584 (2012)}.

\bibitem{Lozovik2012} 
Y. E. Lozovik, S. L. Ogarkov, A. A. Sokolik,
Condensation of electron-hole pairs in a two-layer graphene system: Correlation effects,
\href{https://journals.aps.org/prb/abstract/10.1103/PhysRevB.86.045429}{Phys. Rev. B {\bf 86}, 045429 (2012)}.

\bibitem{Fogler2014} M. M. Fogler, L. V. Butov, K. S. Novoselov,
High-temperature superfluidity with indirect excitons in van der Waals heterostructures,
\href{https://www.nature.com/articles/ncomms5555}{Nature Comm. {\bf 5}, 4555 (2014)}.


\bibitem{Kasprzak2006} 
J. Kasprzak, M. Richard, S. Kundermann, A. Baas, P. Jeambrun, J. M. J. Keeling, F. M. Marchetti, M. H. Szymanska, R. Andre, J. L. Staehli, V. Savona, P. B. Littlewood, B. Deveaud, and Le Si Dang, 
Bose–Einstein condensation of exciton polaritons,
\href{https://www.nature.com/articles/nature05131}{Nature (London) {\bf 443}, 409 (2006)}.

\bibitem{Balili2007} R. Balili, V. Hartwell, D. Snoke, and K. West, 
Bose-Einstein Condensation of Microcavity Polaritons in a Trap,
\href{https://www.science.org/doi/10.1126/science.1140990}{Science {\bf 316}, 1007 (2007)}.

\bibitem{Amo2009} A. Amo, D. Sanvitto, F. P. Laussy, D. Ballarini, E. del Valle, M. D. Martin, A. Lemaitre, J. Bloch, D. N. Krizhanovskii, M. S. Skolnick, C. Tejedor, and L. Vina, 
Collective fluid dynamics of a polariton condensate in a semiconductor microcavity,
\href{https://www.nature.com/articles/nature07640}{Nature {\bf 457}, 291 (2009)}.

\bibitem{Lagoudakis2008} K. G. Lagoudakis, M. Wouters, M. Richard, A. Baas, I. Carusotto, R. Andre, L. S. Dang, and B. Deveaud-Pledran, 
Quantized vortices in an exciton–polariton condensate,
\href{https://www.nature.com/articles/nphys1051}{Nature Phys. {\bf 4}, 706 (2008)}.

\bibitem{Tosi2012} G. Tosi, G. Christmann, N. G. Berloff, P. Tsotsis, T. Gao, Z. Hatzopoulos, P. G. Savvidis, and J. J. Baumberg, 
Geometrically locked vortex lattices in semiconductor quantum fluids,
\href{https://www.nature.com/articles/ncomms2255}{Nat. Commun. {\bf 3}, 1243 (2012)}.

\bibitem{Gao2018} T. Gao, O. A. Egorov, E. Estrecho, K. Winkler, M. Kamp, C. Schneider, S. Hofling, A. G. Truscott, and E. A. Ostrovskaya, 
Controlled Ordering of Topological Charges in an Exciton-Polariton Chain,
\href{https://journals.aps.org/prl/abstract/10.1103/PhysRevLett.121.225302}{Phys. Rev. Lett. {\bf 121}, 225302 (2018)}.

\bibitem{Kwon2019} M.-S. Kwon, B. Y. Oh, S.-H. Gong, J.-H. Kim, H. K. Kang, S. Kang, J. D. Song, H. Choi, and Y.-H. Cho, Direct Transfer of Light’s Orbital Angular Momentum onto a Nonresonantly Excited Polariton Superfluid,
\href{https://journals.aps.org/prl/abstract/10.1103/PhysRevLett.122.045302}{Phys. Rev. Lett. {\bf 122}, 045302 (2019)}.

\bibitem{seki2012} S. Seki, X.Z. Yu, S. Ishiwata, and Y. Tokura,
Observation of Skyrmions in a Multiferroic Material,
\href{https://www.science.org/doi/10.1126/science.1214143}{Science {\bf 336}, 198 (2012)}. 

\bibitem{petrovic2021} A. P. Petrovi\'c, M. Raju, X. Y. Tee, A. Louat, I. Maggio-Aprile, R. M. Menezes, M. J. Wyszy\'nski, N. K. Duong, M. Reznikov, Ch. Renner, M. V. Milo\v{s}evi\'c, and C. Panagopoulos,
Skyrmion-(Anti)Vortex Coupling in a Chiral Magnet-Superconductor Heterostructure,
\href{https://journals.aps.org/prl/abstract/10.1103/PhysRevLett.126.117205}{Phys. Rev. Lett. {\bf 126}, 117205 (2021)}.









\bibitem{Burch2018} K. S. Burch, D. Mandrus, and J.-G. Park, 
Magnetism in two-dimensional van der Waals materials,
\href{https://www.nature.com/articles/s41586-018-0631-z}{Nature {\bf 563}, 47 (2018)}.

\bibitem{Wildes2015} A. R. Wildes, V. Simonet, E. Ressouche, G. J. McIntyre, M. Avdeev, E. Suard, S. A. J. Kimber, D. Lançon, G. Pepe, B. Moubaraki, and T. J. Hicks, 
Magnetic structure of the quasi-two-dimensional antiferromagnet 
NiPS3,
\href{https://journals.aps.org/prb/abstract/10.1103/PhysRevB.92.224408}{Phys. Rev. B {\bf 92}, 224408 (2015)}.

\bibitem{Kang2020} S. Kang, K. Kim, B. H. Kim, J. Kim, K. I. Sim, J.-U. Lee, S. Lee, K. Park, S. Yun, T. Kim, A. Nag, A. Walters, M. Garcia-Fernandez, J. Li, L. Chapon, K.-J. Zhou, Y.-W. Son, J. H. Kim, H. Cheong, and J.-G. Park, 
Coherent many-body exciton in van der Waals antiferromagnet NiPS3,
\href{https://www.nature.com/articles/s41586-020-2520-5}{Nature {\bf 583}, 785 (2020)}.

\bibitem{Ho2021} C.-H. Ho, T.-Y. Hsu, and L. C. Muhimmah,
The band-edge excitons observed in few-layer NiPS3,
\href{https://www.nature.com/articles/s41699-020-00188-8}{NPJ 2D Mater. Appl. {\bf 5}, 1 (2021)}.

\bibitem{Birowska2021} M. Birowska, P. E. Faria Junior, J. Fabian, and J. Kunstmann, 
Large exciton binding energies in  MnPS3 as a case study of a van der Waals layered magnet,
\href{https://journals.aps.org/prb/abstract/10.1103/PhysRevB.103.L121108}{Phys. Rev. B {\bf 103}, L121108 (2021)}.

\bibitem{Huang2017} B. Huang, G. Clark, E. Navarro-Moratalla, D. R. Klein, R. Cheng, K. L. Seyler, D. Zhong, E. Schmidgall, M. A. McGuire, D. H. Cobden, W. Yao, D. Xiao, P. Jarillo-Herrero, and X. Xu,
Layer-dependent ferromagnetism in a van der Waals crystal down to the monolayer limit,
\href{https://www.nature.com/articles/nature22391}{Nature {\bf 546}, 270 (2017)}.

\bibitem{Zheng2018} F. Zheng, J. Zhao, Z. Liu, M. Li, M. Zhou, S. Zhang, and  P. Zhang,
Tunable spin states in the two-dimensional magnet CrI3,
\href{https://pubs.rsc.org/en/content/articlelanding/2018/nr/c8nr03230k}{Nanoscale {\bf 10}, 14298 (2018)}.

\bibitem{Kashin2020} I. V. Kashin, V. V. Mazurenko, M. I. Katsnelson, and A. N. Rudenko, 
Orbitally-resolved ferromagnetism of monolayer CrI3,
\href{https://iopscience.iop.org/article/10.1088/2053-1583/ab72d8}{2D Mater. {\bf 7}, 025036 (2020)}.

\bibitem{kim2019} M. Kim, P. Kumaravadivel, J. Birkbeck, W. Kuang, S. G. Xu, D. G. Hopkinson, J. Knolle, P. A. McClarty, A. I. Berdyugin, M. Ben Shalom, R. V. Gorbachev, S. J. Haigh, S. Liu, J.H. Edgar, K. S. Novoselov, I. V. Grigorieva, and A. K. Geim, 
Micromagnetometry of two-dimensional ferromagnets,
\href{https://www.nature.com/articles/s41928-019-0302-6}{Nat. Electron. {\bf 2}, 457 (2019)}. 

\bibitem{Mermin1966} N. D. Mermin, and H. Wagner,
Absence of ferromagnetism or antiferromagnetism in one- or two-dimensional isotropic Heisenberg models,
\href{https://journals.aps.org/prl/abstract/10.1103/PhysRevLett.17.1133}{Phys. Rev. Lett. {\bf 17}, 1133 (1966)}.

\bibitem{Lado2017} J. L. Lado, and J. Fern\'andez-Rossier,
On the origin of magnetic anisotropy in two dimensional CrI3, 
\href{https://iopscience.iop.org/article/10.1088/2053-1583/aa75ed}{2D Mater. {\bf 4}, 035002 (2017)}.

\bibitem{Liu2018} J. Liu, M. Shi, J. Lu, and M.P. Anantram,
Analysis of electrical-field-dependent Dzyaloshinskii-Moriya interaction and magnetocrystalline anisotropy in a two-dimensional ferromagnetic monolayer, 
\href{https://journals.aps.org/prb/abstract/10.1103/PhysRevB.97.054416}{
Phys. Rev. B {\bf 97}, 054416 (2018)}.


\bibitem{Ghosh2019} S. Ghosh, N. Stoji\'c, and N. Binggeli, 
Structural and magnetic response of CrI3 monolayer to electric field, 
\href{https://www.sciencedirect.com/science/article/pii/S0921452619304065}{
Physica B Condens. Matter {\bf 570}, 166 (2019)}.

\bibitem{Vishkayi2020}
S. I. Vishkayi, Z. Torbatian, A. Qaiumzadeh, and R. Asgari, 
Strain and electric-field control of spin-spin interactions in monolayer CrI3, 
\href{https://journals.aps.org/prmaterials/abstract/10.1103/PhysRevMaterials.4.094004}{Phys. Rev. Mater. {\bf 4}, 094004 (2020)}.

\bibitem{Liu2018a} J. Liu, M. Shi, P. Mo, and J. Lu, 
Electrical-field-induced magnetic Skyrmion ground state in a two-dimensional chromium tri-iodide ferromagnetic monolayer, 
\href{https://aip.scitation.org/doi/10.1063/1.5030441}{AIP Advances {\bf 8}, 055316 (2018)}.

\bibitem{Behera2019} A. K. Behera, S. Chowdhury, and S. R. Das, 
Magnetic skyrmions in atomic thin CrI3 monolayer, 
\href{https://aip.scitation.org/doi/10.1063/1.5096782}{Appl. Phys. Lett. {\bf 114}, 232402 (2019)}.

\bibitem{Seyler2018} K. L. Seyler, D. Zhong, D. R. Klein, S. Gao, X. Zhang, B. Huang, E. Navarro-Moratalla, L. Yang, D. H. Cobden, M. A. McGuire, W. Yao, D. Xiao, P. Jarillo-Herrero, and Xiaodong Xu, 
Ligand-field helical luminescence in a 2D ferromagnetic insulator,
\href{https://www.nature.com/articles/s41567-017-0006-7}{Nat. Phys. {\bf 14}, 277 (2018)}.



\bibitem{Pervishko2020} 
A. A. Pervishko, D. Yudin, V. Kumar Gudelli, A. Delin, O. Eriksson, and G.-Y. Guo, 
Localized surface electromagnetic waves in CrI3-based magnetophotonic structures,
\href{https://opg.optica.org/oe/fulltext.cfm?uri=oe-28-20-29155&id=439736}{Optics Express {\bf 28}, 20 (2020)}.

\bibitem{Kudlis2021} A. Kudlis, I. Iorsh, and I.A. Shelykh, 
All-optical resonant magnetization switching in CrI3 monolayers,
\href{https://journals.aps.org/prb/abstract/10.1103/PhysRevB.104.L020412}{Phys. Rev. B {\bf 104}, L020412 (2021)}.

\bibitem{Kozin2021}
V. K. Kozin, V. A. Shabashov, A. V. Kavokin, and I. A. Shelykh,
Anomalous Exciton Hall Effect,
\href{https://journals.aps.org/prl/abstract/10.1103/PhysRevLett.126.036801}{Phys. Rev. Lett. {\bf 126}, 036801 (2021)}.

\bibitem{Wu2019} 
M. Wu, Z. Li, T. Cao, and S. G. Louie, 
Physical origin of giant excitonic and magnetooptical responses in two-dimensional ferromagnetic insulators,
\href{https://www.nature.com/articles/s41467-019-10325-7}{Nature Comm. {\bf 10}, 2371 (2019)}.


\bibitem{ivanov2021} A. V. Ivanov, V. M. Uzdin, and H. J\'onsson, Fast and robust algorithm for energy minimization of spin systems applied in an analysis of high temperature spin configurations in terms of skyrmion density, \href{https://doi.org/10.1016/j.cpc.2020.107749}{Comput. Phys. Commun. {\bf 260}, 107749 (2021)}.

\bibitem{braun1994}
H.-B. Braun, Fluctuations and instabilities of ferromagnetic domain-wall pairs in an external magnetic field,
\href{https://doi.org/10.1103/PhysRevB.50.16485}{Phys. Rev. B {\bf  50}, 16485 (1994)}.




\bibitem{bogdanov1994a} A. Bogdanov, A. Hubert, Thermodynamically stable magnetic vortex states in magnetic crystals, \href{https://doi.org/10.1016/0304-8853(94)90046-9}{J. Magn. Magn. Mater. \textbf{138} 255-269 (1994)}. 

\bibitem{bogdanov1994b} A. Bogdanov, A. Hubert, Thermodynamically stable magnetic vortex states in magnetic crystals, \href{https://doi.org/10.1002/pssb.2221860223}{Phys. Status Solidi (b) \textbf{186} 527-543 (1994)}.

\bibitem{nagaosa2013}
N. Nagaosa and Y. Tokura, Topological properties and dynamics of magnetic skyrmions, 
\href{https://doi.org/10.1038/nnano.2013.243}{Nat. Nanotechnol. {\bf  8}, 899-911 (2013)}.

\bibitem{Denisov2016}
K. S. Denisov, I. V. Rozhansky, N. S. Averkiev, and E. Lähderanta,
Electron Scattering on a Magnetic Skyrmion in the Nonadiabatic Approximation,
\href{https://journals.aps.org/prl/abstract/10.1103/PhysRevLett.117.027202}{Phys. Rev. Lett. {\bf 117}, 027202 (2016)}.

\bibitem{Denisov2020}
K. S. Denisov, 
Theory of an electron asymmetric scattering on skyrmion textures in two-dimensional systems,
\href{https://iopscience.iop.org/article/10.1088/1361-648X/ab966e}{J. Phys.: Condens. Matter {\bf 32}, 415302 (2020)}.
 
\bibitem{Dalibard2011}
J. Dalibard, F. Gerbier, G. Juzeliūnas, and P. Öhberg,
Colloquium: Artificial gauge potentials for neutral atoms,
\href{https://journals.aps.org/rmp/abstract/10.1103/RevModPhys.83.1523}{Rev. Mod. Phys. {\bf 83}, 1523 (2011)}.

\bibitem{Ritz2013} 
R. Ritz, M. Halder, C. Franz, A. Bauer, M. Wagner, R. Bamler, A. Rosch, and C. Pfleiderer,
Giant generic topological Hall resistivity of MnSi under pressure,
\href{https://journals.aps.org/prb/abstract/10.1103/PhysRevB.87.134424}{Phys. Rev. B {\bf 87}, 134424 (2013)}.

\bibitem{Sinitsyn2007} 
N. A. Sinitsyn, A. H. MacDonald, T. Jungwirth, V. K. Dugaev, and J. Sinova,
Anomalous Hall effect in a two-dimensional Dirac band: The link between the Kubo-Streda
formula and the semiclassical Boltzmann equation approach,
\href{https://journals.aps.org/prb/abstract/10.1103/PhysRevB.75.045315}{Phys. Rev. B {\bf 75}, 045315 (2007)}.

\bibitem{Aharonov1959}
Y. Aharonov, and D. Bohm,
Significance of Electromagnetic Potentials in the Quantum Theory,
\href{https://journals.aps.org/pr/abstract/10.1103/PhysRev.115.485}{Phys. Rev. {\bf 115}, 485 (1959)}. 

\bibitem{Mortensen2005} 
J. J. Mortensen and L. B. Hansen and K. W. Jacobsen, 
Real-space grid implementation of the projector augmented wave method,
\href{https://journals.aps.org/prb/abstract/10.1103/PhysRevB.71.035109}{Phys. Rev. B {\bf 71}, 035109 (2005)}.

\bibitem{Enkovaara2010} J. Enkovaara, C. Rostgaard, J. J. Mortensen, J. Chen, M. Dułak, L. Ferrighi, J. Gavnholt, C. Glinsvad, V. Haikola, H. A. Hansen, H. H. Kristoffersen, M. Kuisma, A. H. Larsen, L. Lehtovaara, M. Ljungberg, O. Lopez-Acevedo, P. G. Moses, J. Ojanen, T. Olsen, V. Petzold, N. A. Romero, J. Stausholm-Møller, M. Strange, G. A. Tritsaris, M. Vanin, M. Walter, B. Hammer, H. Häkkinen, G. K. H. Madsen, R. M. Nieminen, J. K. Nørskov, M. Puska, T. T. Rantala, J. Schiøtz, K. S. Thygesen, and K. W. Jacobsen,
Electronic structure calculations with GPAW: a real-space implementation of the projector augmented-wave method,
\href{https://iopscience.iop.org/article/10.1088/0953-8984/22/25/253202}{J. Phys.: Condens. Matter {\bf 22}, 253202 (2010)}.

\bibitem{Yan2011} J. Yan, J. J. Mortensen, K. W. Jacobsen, and K. S. Thygesen,
Linear density response function in the projector augmented wave method: Applications to solids, surfaces, and interfaces,
\href{https://journals.aps.org/prb/abstract/10.1103/PhysRevB.83.245122}{Phys. Rev. B {\bf 83}, 245122 (2011)}.


\bibitem{Olsen2016} T. Olsen, 
Designing in-plane heterostructures of quantum spin Hall insulators from first principles: 1T'-MoS2 with adsorbates,
\href{https://journals.aps.org/prb/abstract/10.1103/PhysRevB.94.235106}{Phys. Rev. B {\bf 94}, 235106 (2016)}.

\bibitem{Olsen2021} T. Olsen, 
Unified Treatment of Magnons and Excitons in Monolayer CrI$_3$ from Many-Body Perturbation Theory, 
\href{https://doi.org/10.1103/PhysRevLett.127.166402}{Phys. Rev. Lett. {\bf 127}, 166402 (2021)}.

\bibitem{Huser2013} F. Huser, T. Olsen and K. S. Thygesen,
How dielectric screening in two-dimensional crystals affects the convergence of excited-state calculations: Monolayer MoS2,
\href{https://journals.aps.org/prb/abstract/10.1103/PhysRevB.88.245309}{Phys. Rev. B {\bf 88}, 245309 (2013)}.

\bibitem{Levine1989}
Z. Levine, and D. Allan,
Linear optical response in silicon and germanium including self-energy effects. 
\href{https://link.aps.org/doi/10.1103/PhysRevLett.63.1719}{ Phys. Rev. Lett. {\bf 63}, 1719-1722 (1989)}. 

\bibitem{Yang2015} H. Yang, A. Thiaville, S. Rohart, A. Fert, and M. Chshiev,
Anatomy of Dzyaloshinskii-Moriya Interaction at Co/Pt Interfaces,
\href{https://journals.aps.org/prl/abstract/10.1103/PhysRevLett.115.267210}{Phys. Rev. Lett. {\bf 115}, 267210 (2015)}.









\bibitem{Adhikari1986}  
S. K. Adhikari,
Quantum scattering in two dimensions,
\href{https://aapt.scitation.org/doi/10.1119/1.14623}{ Am. J. Phys. {\bf 54}, 362 (1986)}.





\end{thebibliography}
\end{document}